\documentclass[sigconf,]{acmart}

\usepackage{bm}
\usepackage{amsmath}
\usepackage{booktabs}
\usepackage{multirow}
\usepackage{multicol}
\usepackage[normalem]{ulem}
\useunder{\uline}{\ul}{}
\usepackage{appendix}
\usepackage{balance}
\usepackage{makecell}
\usepackage{color}
\usepackage{colortbl}
\usepackage[marginal]{footmisc}
\usepackage[linesnumbered,ruled,vlined, noend]{algorithm2e}
\usepackage{xcolor}
\usepackage{subcaption}
\usepackage[most]{tcolorbox}
\usepackage{xcolor}

\definecolor{casegreenbg}{RGB}{240,247,236}

\definecolor{casegreenbase}{RGB}{84,130,53}

\colorlet{casegreen}{casegreenbase!80!white}

\definecolor{caseorangebg}{RGB}{254,249,245}

\definecolor{caseorange}{RGB}{197,90,17}

\definecolor{caseblue}{HTML}{4472C4}
\definecolor{casered}{HTML}{C00000}

\newcommand{\casepos}[1]{\textcolor{caseblue}{#1}}
\newcommand{\caseneg}[1]{\textcolor{casered}{#1}}

\newtcolorbox{successcasebox}[1]{
    enhanced,
    breakable,
    width=\columnwidth,
    colback=casegreenbg,
    colframe=casegreen,
    colbacktitle=casegreen,
    coltitle=white,
    title={#1},
    fonttitle=\bfseries\small,
    fontupper=\small,
    boxrule=1pt,
    arc=2mm,
    outer arc=2mm,
    left=3mm,
    right=3mm,
    top=2mm,
    bottom=2mm,
    boxsep=0pt,
    before skip=7pt,
    after skip=6pt
}

\newtcolorbox{failurecasebox}[1]{
    enhanced,
    breakable,
    width=\columnwidth,
    colback=caseorangebg,
    colframe=caseorange,
    colbacktitle=caseorange,
    coltitle=white,
    title={#1},
    fonttitle=\bfseries\small,
    fontupper=\small,
    boxrule=1pt,
    arc=2mm,
    outer arc=2mm,
    left=3mm,
    right=3mm,
    top=2mm,
    bottom=2mm,
    boxsep=0pt,
    before skip=7pt,
    after skip=6pt
}

\definecolor{bgcolor}{RGB}{242, 242, 242}
\usepackage{varwidth}
\usepackage{caption}
\usepackage{graphicx}
\usepackage{float} 
\usepackage{algorithmic}
\usepackage{marvosym}

\usepackage{enumitem}
\usepackage{subcaption}
\usepackage{graphicx}
\graphicspath{{./}{upload/}}

\setitemize[1]{topsep=0pt}

\AtBeginDocument{%
  }

\setcopyright{acmlicensed}
\copyrightyear{2018}
\acmYear{2018}
\acmDOI{XXXXXXX.XXXXXXX}
\acmConference[Conference acronym 'XX]{Make sure to enter the correct
  conference title from your rights confirmation email}{June 03--05,
  2018}{Woodstock, NY}
\acmISBN{978-1-4503-XXXX-X/2018/06}

\begin{document}

\title{Personalized Communication Skills for Agentic Recommender Systems}

\author{Zongwei Wang}
\email{zongwei@cqu.edu.cn}
\orcid{0000-0002-9774-4596}
\affiliation{%
  \institution{Chongqing University}
  \city{Chongqing}
  \country{China}
}
  
\author{Min Gao}
\email{gaomin@cqu.edu.cn}
\authornote{Corresponding author}
\affiliation{%
  \institution{Chongqing University}
  \city{Chongqing}
  \country{China}}

\author{Guangyu Hu}
\email{huguangyu@stu.cqu.edu.cn}
\orcid{0009-0006-2402-4287}
\affiliation{%
  \institution{Chongqing University}
  \city{Chongqing}
  \country{China}
}

\author{Xinyi Gao}
\email{xinyi.gao@uq.edu.au}
\affiliation{%
  \institution{The University of Queensland}
  \city{Brisbane}
  \country{Australia}}

\author{Junliang Yu}
\email{jl.yu@uq.edu.au}
\affiliation{%
  \institution{The University of Queensland}
  \city{Brisbane}
  \country{Australia}}

\renewcommand{\shortauthors}{Trovato et al.}

\begin{abstract}
Agentic recommender systems increasingly employ large language model-based UserAgents to evaluate candidate items through simulated feedback before recommendations are delivered. However, existing UserAgents typically reason in isolation based on limited personal histories, which may lead to \textit{perspective narrowing}: the agent evaluates candidates from a local and incomplete view, overlooks relevant preference facets, and consequently produces inaccurate judgments. A natural way to alleviate this problem is to introduce other users as advisor agents, whose diverse histories provide complementary evidence that helps the target user reconsider overlooked preference signals. Nevertheless, a generic user-advisor communication process is insufficient, as different user decision states require different forms of external advice. Based on this insight, we propose AgentCom, a personalized communication skill framework for agentic recommender systems. AgentCom organizes reusable communication skills into a shared \textit{why--what--how--who} skill bank: \textit{why} identifies the decision deficiency that necessitates communication, \textit{what} specifies the information task, \textit{how} determines the advisor interaction protocol, and \textit{who} retrieves advisors capable of executing that protocol. To make the shared skill bank personalized at use time and adaptive over time, AgentCom introduces two complementary mechanisms: personalized skill routing and failure-driven skill evolution. Personalized skill routing constructs a communication path by sequentially selecting suitable skills for each user and recommendation context. Failure-driven skill evolution learns from unsuccessful communication cases and enriches the shared bank with reusable skills that address previously uncovered communication needs. Experiments on multiple benchmarks show that AgentCom consistently improves recommendation performance across traditional, social, and agentic recommenders. 
\end{abstract}



\keywords{Large Language Model, Agent, Skill, Recommender Systems}


\received{20 February 2007}
\received[revised]{12 March 2009}
\received[accepted]{5 June 2009}

\maketitle

\section{Introduction}

Recommender systems have traditionally followed a one-shot prediction paradigm, in which a model infers user preferences from past behaviors and directly produces a ranked list of items~\cite{01kang2018self,21wang2023efficient}. While this paradigm has achieved widespread success, it treats recommendation as a static mapping and provides limited opportunities to verify whether the inferred preferences are accurate after the results are generated. Recent advances in large language models (LLM) enable autonomous agents to simulate user responses~\cite{27wang2025user,28wang2024recmind}, introducing a new paradigm in which recommendations can be evaluated before delivery. In this setting, the UserAgents assess candidate items according to the user's profile and interaction history, and their feedback guides the recommender to revise the results, helping identify preference mismatches and improve recommendation quality~\cite{13zhang2024agentcf,14liu2025agentcf++,15xia2025multi}.

\begin{figure}[t]
\centering

\begin{subfigure}[t]{0.48\textwidth}
\centering
\includegraphics[width=\linewidth]{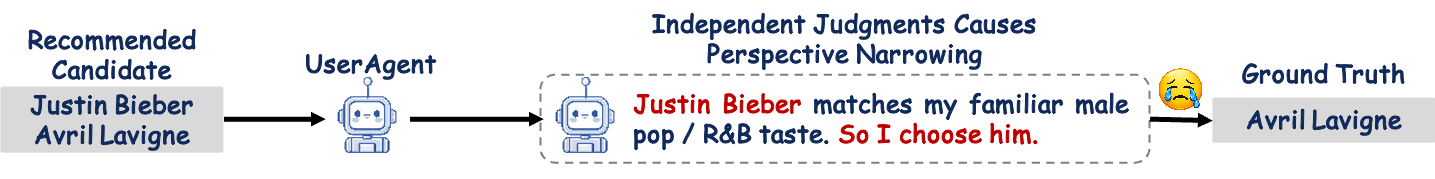}
\caption{Independent judgment paradigm.}
\label{fig:independent_judgment}
\end{subfigure}

\vspace{0.6em}

\begin{subfigure}[t]{0.48\textwidth}
\centering
\includegraphics[width=\linewidth]{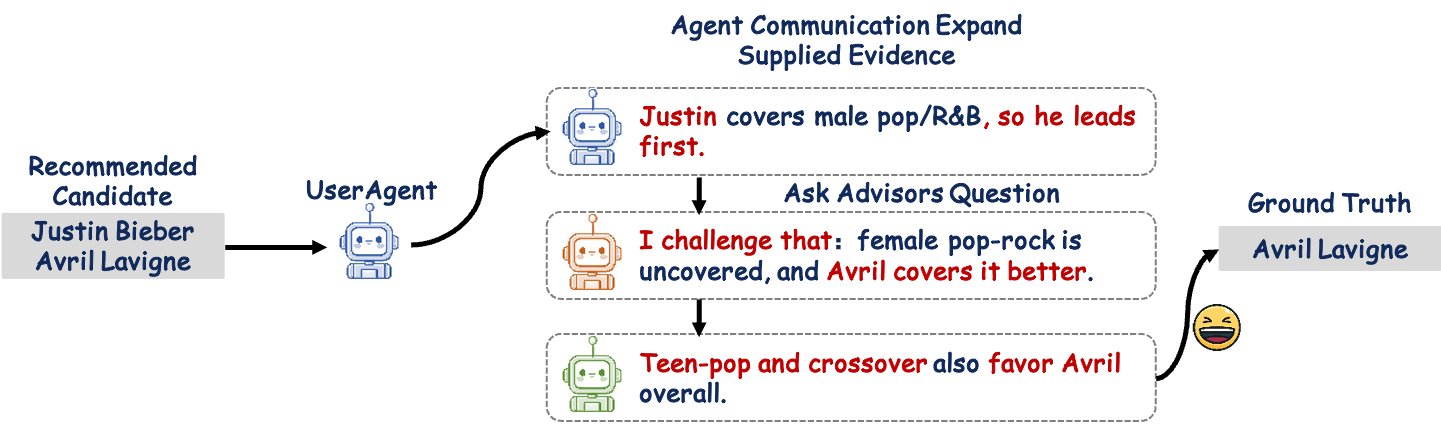}
\caption{Advisor-assisted agent communication paradigm.}
\label{fig:advisor_communication}
\end{subfigure}

\caption{Comparison between independent judgment and advisor-assisted 
judgment through agent communication.}
\label{fig:communication_motivation}
\end{figure}

Nevertheless, existing UserAgents typically make such judgments independently, 
relying primarily on their own interaction histories and internally generated 
reasoning~\cite{03chen2026memrec,17li2026recnet}. This isolated process may lead to \textit{perspective narrowing}~\cite{36liang2024encouraging,37du2024improving}: 
the agent evaluates candidates from a local and incomplete view constrained by 
limited personal evidence. Figure~\ref{fig:communication_motivation} illustrates 
this limitation and how agent communication can alleviate it. As shown in 
Figure~\ref{fig:communication_motivation}(a), the target UserAgent relies on familiar 
male pop/R\&B evidence and incorrectly selects Justin Bieber while overlooking 
other relevant preference facets. In contrast, the advisor agents in 
Figure~\ref{fig:communication_motivation}(b) introduce complementary evidence about 
female pop-rock, teen-pop, and crossover preferences, providing support that is 
unavailable in the target user's independent reasoning. Consequently, advisor communication enables the target UserAgent to reconsider 
its initial judgment by combining its own preferences with evidence supplied by 
advisor agents, rather than simply preserving existing biases. The effectiveness 
of such advisor-assisted judgment is also supported by studies of human advice 
taking~\cite{34yaniv2004receiving,35yaniv2007using}.


However, directly applying the same communication process across different decision states may be suboptimal, as communication is useful for different reasons in different decision states. A cold-start user may need evidence that compensates for a sparse history~\cite{39ma2009learning}, a hesitant user may need help distinguishing several plausible candidates~\cite{40tversky1992choice}, and a user facing an unfamiliar item may need evidence about novelty and risk~\cite{41lewis2004knowledge}. These needs suggest that effective communication should not follow a fixed procedure, but instead be composed from reusable communication skills that can be selectively invoked and combined according to the current decision state. Once the system identifies \textit{why} communication is needed, it should determine \textit{what} information is required, \textit{how} the discussion should be organized, and \textit{who} can execute the selected protocol. Inspired by Lasswell's classical communication model~\cite{38lasswell1948structure}, we decompose agent communication into four observable decision layers: \textit{why}, which identifies the decision deficiency that makes external advice necessary; \textit{what}, which converts that deficiency into a concrete information task; \textit{how}, which specifies the advisor interaction protocol and thereby determines the required communication shape; and \textit{who}, which retrieves advisors with relevant or complementary evidence. 

Building on these insights, we propose AgentCom, 
a personalized \underline{\textbf{Com}}munication skill framework for
\underline{\textbf{Agent}}ic recommender systems. AgentCom maintains a shared hierarchical communication skill bank organized by the \textit{why--what--how--who} dimensions. Each user accesses the bank through personalized skill routing that selects a communication path according to the current decision state and available advisor evidence. Because an initialized skill bank cannot anticipate every communication need, 
AgentCom analyzes failed decisions to locate the responsible layer and determine 
why the current communication failed. It then either refines an existing skill 
for a more specific case or introduces a new skill for a communication strategy 
not covered by the bank. The affected users are subsequently rerouted over the 
evolved bank. Our contributions can be summarized as follows:
\begin{itemize}[leftmargin=*]
\item To the best of our knowledge, we are the first to introduce agent communication as personalized skills into agentic recommender systems, enabling each UserAgent to acquire and execute communication strategies tailored to its decision state.
\item We propose AgentCom, which organizes reusable capabilities into a 
hierarchical \textit{why--what--how--who} skill bank, constructs personalized 
routes over shared skills, and evolves the bank from failed decisions through 
skill refinement and generation.
\item Experiments on three datasets show that AgentCom consistently improves recommendation performance across traditional, social, and agentic recommenders, demonstrating the importance of agent communication in recommendation.
\end{itemize}

\section{Preliminary}
\label{Preliminary}

We first formulate the independent judgment process used by existing 
UserAgents over a recommender-provided candidate set. We then introduce an 
advisor-assisted extension that preserves the same judgment interface while 
allowing the target UserAgent to revise its initial decision using external 
evidence from other users.

\subsection{Independent UserAgent Judgment}
\label{sec:single_user_preliminary}

Let $\mathcal{U}$ and $\mathcal{I}$ denote the sets of users and items, 
respectively. For each user $u\in\mathcal{U}$, let $h_u$ denote the historical 
interaction sequence. At recommendation turn $t$, a base recommender produces 
a candidate set $\mathcal{C}_{u,t}\subseteq\mathcal{I}$. The base recommender 
may be a conventional recommendation model or an agentic recommender. Let 
$\mathcal{M}_{u,t-1}$ denote the interaction memory accumulated before the 
current turn. Given these inputs, the UserAgent independently evaluates the candidate set and 
forms an initial judgment:
\begin{equation}
\begin{aligned}
\mathbf{q}^{(0)}_{u,t}
=
\left(
i^{(0)}_{u,t},
d^{(0)}_{u,t},
j^{(0)}_{u,t},
\mathcal{H}^{(0)}_{u,t}
\right)
\sim
\pi_{\mathrm{user}}
\left(
\cdot
\mid
h_u,
\mathcal{C}_{u,t},
\mathcal{M}_{u,t-1}
\right),
\end{aligned}
\label{eq:provisional_user_judgment}
\end{equation}
where $\pi_{\mathrm{user}}$ denotes the UserAgent policy,
$i^{(0)}_{u,t}\in\mathcal{C}_{u,t}$ is the item that the UserAgent prefers most among all candidates in $\mathcal{C}_{u,t}$, and $j^{(0)}_{u,t}$ is the rationale supporting this choice.
The variable $d^{(0)}_{u,t}\in\{0,1\}$ represents the decision status:
$d^{(0)}_{u,t}=1$ indicates that the UserAgent is willing to commit to its
current choice, whereas $d^{(0)}_{u,t}=0$ indicates that the judgment remains
unresolved. The hesitation set
$\mathcal{H}^{(0)}_{u,t}\subseteq
\mathcal{C}_{u,t}\setminus\{i^{(0)}_{u,t}\}$
contains plausible alternatives that the UserAgent cannot reliably distinguish
from its current choice.

Without advisor communication, the initial judgment directly serves as the final judgment. In this case, the UserAgent directly returns its selected item, decision status, and supporting rationale, while the hesitation set $\mathcal{H}^{(0)}_{u,t}$ is not used for any subsequent communication or judgment revision:
\begin{equation}
\mathbf{m}^{\mathrm{ind}}_{u,t}
=
\left(
i^{(0)}_{u,t},
d^{(0)}_{u,t},
j^{(0)}_{u,t}
\right),
\label{eq:independent_user_output}
\end{equation}
and the judgment is incorporated into the interaction memory:
\begin{equation}
\mathcal{M}_{u,t}
=
\operatorname{Update}
\left(
\mathcal{M}_{u,t-1},
\mathcal{C}_{u,t},
\mathbf{q}^{(0)}_{u,t}
\right).
\label{eq:independent_user_memory}
\end{equation}

This process allows the UserAgent to examine and potentially revise the item
prioritized by the base recommender. However, the judgment is derived entirely
from the target user's history, previous memory, and the UserAgent's internal
reasoning. The agent therefore lacks external evidence with which to verify,
complement, or challenge its interpretation of the candidates. This isolated
reasoning process may lead to \emph{perspective narrowing}, in which limited
personal evidence constrains the judgment and encourages the UserAgent to
preserve the same local view.

\subsection{Advisor-Assisted UserAgent Judgment}
\label{sec:advisor_assisted_judgment}

Advisor-assisted judgment retains the same candidate set, recommendation input,
and initial UserAgent judgment defined in
Eq.~\eqref{eq:provisional_user_judgment}. The only difference is that
$\mathbf{q}^{(0)}_{u,t}$ is treated as a provisional rather than final
judgment. Before committing to a final choice, the target UserAgent may obtain
complementary evidence from one or more other users.

Let $K_{u,t}$ denote the number of user--advisor communication rounds conducted
for user $u$ at turn $t$. At communication round
$k\in\{1,\ldots,K_{u,t}\}$, the system selects an advisor group according to
the target user's latest judgment:
\begin{equation}
\mathcal{V}^{(k)}_{u,t}
=
\operatorname{SelectAdvisors}
\left(
u,
\mathbf{q}^{(k-1)}_{u,t},
\mathcal{C}_{u,t},
\mathcal{U}\setminus\{u\}
\right),
\label{eq:iterative_advisor_selection}
\end{equation}
where
$\mathcal{V}^{(k)}_{u,t}\subseteq\mathcal{U}\setminus\{u\}$
contains either a single advisor or multiple advisors. A basic selection
policy may rely on random sampling, social relations, preference similarity,
or candidate-related experience.

The selected advisor agents inspect the latest UserAgent judgment using their
own interaction histories and produce complementary evidence:
\begin{equation}
\mathbf{a}^{(k)}_{u,t}
\sim
\pi_{\mathrm{adv}}
\left(
\cdot
\mid
\left\{
h_v
\right\}_{v\in\mathcal{V}^{(k)}_{u,t}},
\mathbf{q}^{(k-1)}_{u,t},
\mathcal{C}_{u,t}
\right),
\label{eq:iterative_advisor_feedback}
\end{equation}
where $\pi_{\mathrm{adv}}$ denotes the advisor policy and
$\mathbf{a}^{(k)}_{u,t}$ contains candidate-level support, objections,
comparisons, or other evidence relevant to the current judgment.

After receiving this evidence, the target UserAgent revises its judgment while
preserving the same output structure as in
Eq.~\eqref{eq:provisional_user_judgment}:
\begin{equation}
\begin{aligned}
\mathbf{q}^{(k)}_{u,t}
&=
\left(
i^{(k)}_{u,t},
d^{(k)}_{u,t},
j^{(k)}_{u,t},
\mathcal{H}^{(k)}_{u,t}
\right)\\
&\sim
\pi_{\mathrm{user}}
\left(
\cdot
\mid
h_u,
\mathcal{C}_{u,t},
\mathcal{M}_{u,t-1},
\mathbf{q}^{(k-1)}_{u,t},
\mathbf{a}^{(k)}_{u,t}
\right).
\end{aligned}
\label{eq:iterative_user_revision}
\end{equation}

The revised choice $i^{(k)}_{u,t}$ may retain or overturn the previous choice,
while $\mathcal{H}^{(k)}_{u,t}$ records the alternatives that remain unresolved
after incorporating the newly obtained advisor evidence.

After $K_{u,t}$ communication rounds, the final judgment is $\mathbf{q}^{(K_{u,t})}_{u,t}$, and the advisor-assisted UserAgent returns:
\begin{equation}
\mathbf{m}^{\mathrm{adv}}_{u,t}
=
\left(
i^{(K_{u,t})}_{u,t},
d^{(K_{u,t})}_{u,t},
j^{(K_{u,t})}_{u,t}
\right).
\label{eq:final_advisor_augmented_evaluation}
\end{equation}

The complete judgment and communication trace is then added to the interaction
memory:
\begin{equation}
\begin{aligned}
\mathcal{M}_{u,t}
=
\operatorname{Update}
\Bigg(
\mathcal{M}_{u,t-1},
\mathcal{C}_{u,t},
\mathbf{q}^{(0)}_{u,t},
\left\{
\left(
\mathcal{V}^{(k)}_{u,t},
\mathbf{a}^{(k)}_{u,t},
\mathbf{q}^{(k)}_{u,t}
\right)
\right\}_{k=1}^{K_{u,t}}
\Bigg).
\end{aligned}
\label{eq:advisor_augmented_memory}
\end{equation}

Nevertheless, merely adding advisors does not specify how useful communication
should be constructed. Without a \textit{why} mechanism, the system cannot
diagnose the deficiency that makes external evidence necessary. Without
\textit{what}, advisors may provide information unrelated to the target
UserAgent's actual uncertainty. Without \textit{how}, the system cannot
determine whether the task requires one advisor or multiple advisors interacting
independently, cooperatively, or competitively. Finally, without a
context-aware \textit{who} mechanism, the system may select advisors who lack
relevant experience or complementary evidence. We therefore introduce
\textbf{AgentCom}, which structures advisor-assisted judgment through reusable
\textit{why--what--how--who} communication skills.

\section{Methodology}
\label{sec:methodology}

\begin{figure*}[t]
\centering
\includegraphics[width=\textwidth]{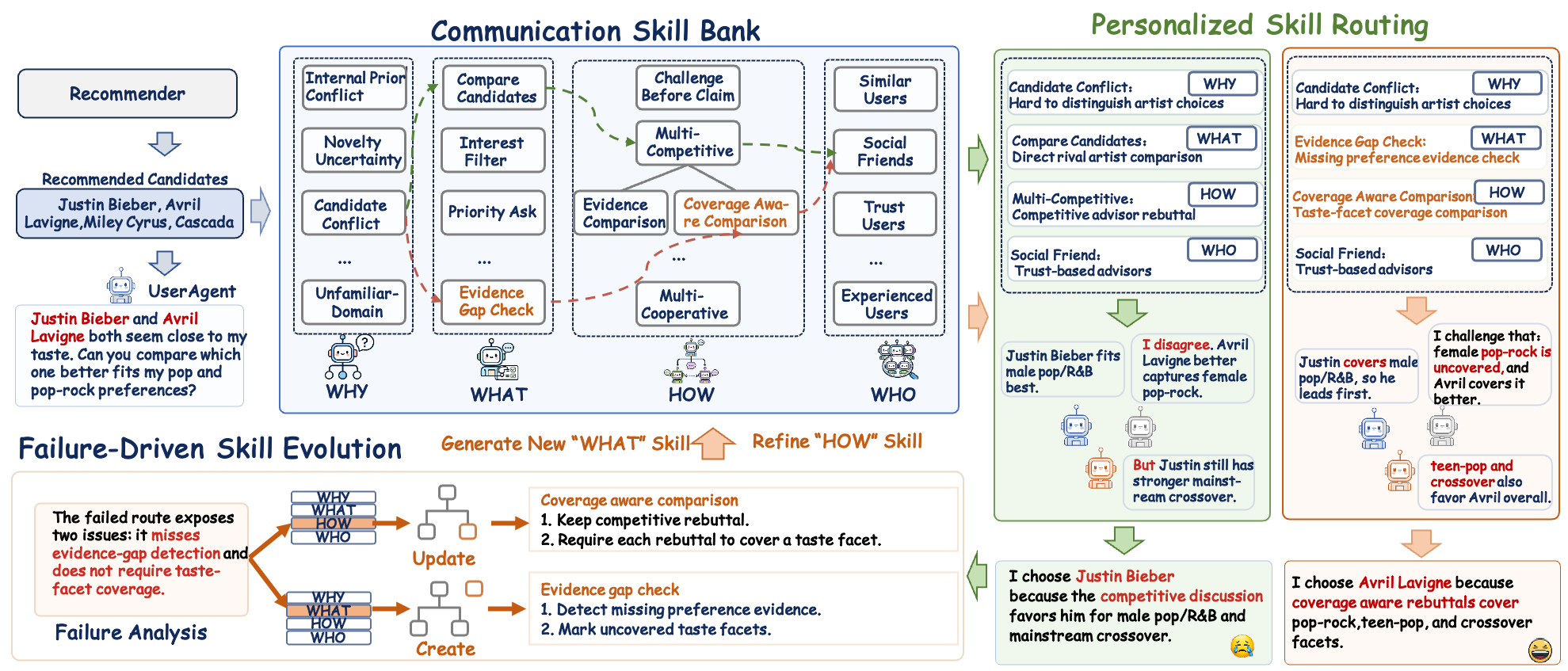}
\caption{Overview of AgentCom. The target UserAgent first evaluates recommender-generated candidates. AgentCom selects a personalized \textit{why--what--how--who} path from the public communication skill bank, executes the corresponding advisor interaction, and returns structured evidence for user re-decision. Failed communication cases are further used to refine personalized routes or expand the public bank.}
\label{fig:framework}
\end{figure*}

AgentCom organizes agent communication as reusable skills rather than constructing a new communication prompt for every user and recommendation. Its central component is a shared public communication skill bank structured along four dimensions: \textit{why}, \textit{what}, \textit{how}, and \textit{who}. For each recommendation, AgentCom selects a complete path through these dimensions, executes the resulting advisor communication, and returns structured evidence to the target UserAgent for re-decision. Figure~\ref{fig:framework} presents the overall framework. A base recommender first generates a candidate set, and the target UserAgent forms an initial judgment. AgentCom then selects a personalized communication path from the public bank and executes the corresponding advisor interaction. If the final decision is unsuccessful, the communication trace is analyzed to identify whether the current route is unsuitable or the bank lacks a reusable capability. The resulting update is incorporated into subsequent routing. We first introduce the communication skill bank in Section~\ref{sec:skill_bank}, then describe the process of AgentCom in Section~\ref{sec:agentcom_process}.

\subsection{Communication Skill Bank}
\label{sec:skill_bank}

AgentCom maintains a hierarchical public bank so that communication capabilities can be reused across users and recommendation contexts. At evolution round $r$, the bank is represented as:
\begin{equation}
\begin{aligned}
\mathcal{B}_{r}=\{&
\mathcal{B}^{\mathrm{why}}_{r},
\mathcal{B}^{\mathrm{what}}_{r},
\mathcal{B}^{\mathrm{how}}_{r},
\mathcal{B}^{\mathrm{who}}_{r}
\},
\end{aligned}
\label{eq:skill_bank}
\end{equation}
where each sub-bank represents one decision in the communication process. 
A complete communication path for user $u$ at turn $t$ is:
\begin{equation}
\begin{aligned}
\mathbf{p}_{u,t}=\big(&
s^{\mathrm{why}}_{u,t},
s^{\mathrm{what}}_{u,t},
s^{\mathrm{how}}_{u,t},
s^{\mathrm{who}}_{u,t}
\big).
\end{aligned}
\label{eq:communication_path}
\end{equation}

To support systematic selection and evolution, each communication skill is formalized as a structured node:
\begin{equation}
s=\left(c_s,a_s,o_s\right),
\label{eq:skill_node}
\end{equation}
where $c_s$ is the applicability condition, $a_s$ is the executable instruction, and $o_s$ is the output contract. The applicability condition describes the observable circumstances where the node is suitable. The executable instruction specifies the operation performed after selection. The output contract specifies the information that the node should return. It enables outputs to be consistently summarized and used by the target UserAgent.

Within each dimension, the nodes are organized hierarchically rather than stored as a flat collection. A parent node represents a general communication capability, while its children specialize that capability for more specific conditions. Nodes sharing the same parent represent alternative strategies for the same communication need. For example, a general multi-advisor capability can contain cooperative and competitive protocols as alternative child nodes. This structure preserves the relationships among skills at different levels of granularity. It also provides a basis for later expansion: a new capability can be introduced as either a specialized child or an alternative sibling without modifying other skills in the bank.

\subsection{AgentCom Process}
\label{sec:agentcom_process}

Given the bank structure above, AgentCom first initializes a compact 
version $\mathcal{B}_0$. 
Each capability 
is instantiated as a structured node following Eq.~\eqref{eq:skill_node}. The 
goal is to provide a reusable starting space that can be subsequently extended, 
rather than to enumerate all possible communication strategies. The complete 
initialization is presented in Appendix~\ref{app:skill_initialization}. Based on 
$\mathcal{B}_0$, AgentCom then proceeds through personalized skill routing and failure-driven skill evolution.

\subsubsection{Personalized Skill Routing}
\label{sec:personalized_routing}

The public bank defines the communication capabilities available to AgentCom, but a separate mechanism is still required to determine how these capabilities should be composed for each user. AgentCom therefore adopts a shared-capability, personalized-routing design: the skill definitions are stored globally in $\mathcal{B}_r$, while their ordering and selection are dynamically adapted to the context of each user.

At turn $t$, the target UserAgent first forms an initial judgment 
$\mathbf{q}^{(0)}_{u,t}$ over the candidate set. The judgment contains its current choice, decision status, supporting rationale, and unresolved alternatives. An LLM-based router analyzes this judgment together with the user's 
history and selects the \textit{why} node that best characterizes the dominant 
deficiency. Every interaction selects a \textit{why} node and proceeds through 
a complete communication path.

The remaining layers are resolved through conditional ordered lists. Let 
$\mathcal{X}_{u,t}$ denote the routing context, including the user history 
$h_u$, the initial judgment $\mathbf{q}^{(0)}_{u,t}$, the candidate set 
$\mathcal{C}_{u,t}$, and the advisor evidence currently available for these 
candidates. Given an upstream node $x$, the LLM router receives 
$\mathcal{X}_{u,t}$ and the descriptions and applicability conditions of all 
active public nodes in the next layer. It then returns these nodes in descending 
order:
\begin{equation}
L^{\ell}_{u,t}[x]
=
\operatorname{Rank}_{\mathrm{LLM}}
\left(
\mathcal{B}^{\ell}_r
\mid x,\mathcal{X}_{u,t}
\right),
\label{eq:llm_conditional_order}
\end{equation}
where $\ell\in\{\mathrm{what},\mathrm{how},\mathrm{who}\}$ and
$\operatorname{Rank}_{\mathrm{LLM}}(\cdot)$ denotes the above LLM-based
analysis and ordering process. The resulting list $L^{\ell}_{u,t}[x]$ places
the most relevant and executable node first, followed by progressively less
suitable alternatives. 

Moreover, the ranking is 
conditional because the selected node in the current layer becomes part of the 
input for ranking the next layer. Specifically, the selected \textit{why} node 
conditions the \textit{what} ranking, the selected \textit{what} node conditions 
the \textit{how} ranking, and the selected \textit{how} node conditions the 
\textit{who} ranking. The routing process is therefore formulated as:
\begin{equation}
\begin{aligned}
s^{\mathrm{what}}_{u,t}
&=
\operatorname{Select}
\left(
L^{\mathrm{what}}_{u,t}
[s^{\mathrm{why}}_{u,t}]
\right),\\
s^{\mathrm{how}}_{u,t}
&=
\operatorname{Select}
\left(
L^{\mathrm{how}}_{u,t}
[s^{\mathrm{what}}_{u,t}]
\right),\\
s^{\mathrm{who}}_{u,t}
&=
\operatorname{Select}
\left(
L^{\mathrm{who}}_{u,t}
[s^{\mathrm{how}}_{u,t}]
\right),
\end{aligned}
\label{eq:conditional_route}
\end{equation}
where $\operatorname{Select}(\cdot)$ traverses an ordered list and returns the 
first node whose applicability condition is satisfied. If a highly ranked node 
cannot be applied, the router continues to the next node rather than 
reconstructing the entire path.

The final selections form the communication path $\mathbf{p}_{u,t}$ in 
Eq.~\eqref{eq:communication_path}. Different users can therefore use the same 
public nodes while following different paths. Once $\mathbf{p}_{u,t}$ is selected, AgentCom converts the four nodes into an executable communication plan. After communication, AgentCom organizes the advisor responses according to the output contracts of the selected nodes. The resulting evidence packet $\mathcal{A}_{u,t}$ records candidate-level support, risks, disagreements, corrections, and remaining evidence gaps. The packet summarizes the discussion without replacing the target UserAgent's authority. The UserAgent combines this external evidence with its initial judgment and personal history to produce the final decision.


\subsubsection{Failure-Driven Skill Evolution}
\label{sec:skill_evolution}

The initialized bank provides a starting point, but it cannot anticipate 
every communication need across different users and recommendation contexts. 
Failure analysis therefore aims not only to identify an incorrect final decision, 
but also to locate the communication layer responsible for the failure and 
determine whether an existing skill 
should be specialized or a new communication capability should be introduced.

During training, let $K_{u,t}$ denote the number of communication and re-decision 
rounds completed by user $u$ at turn $t$. The complete sequence of UserAgent 
judgments is represented as:
\begin{equation}
\mathcal{Q}_{u,t}
=
\left(
\mathbf{q}^{(0)}_{u,t},
\mathbf{q}^{(1)}_{u,t},
\ldots,
\mathbf{q}^{(K_{u,t})}_{u,t}
\right),
\label{eq:decision_trajectory}
\end{equation}
where $\mathbf{q}^{(0)}_{u,t}$ is the initial judgment and 
$\mathbf{q}^{(K_{u,t})}_{u,t}$ is the final judgment after all communication 
rounds. 

Let $i^{(K_{u,t})}_{u,t}$ denote the item selected in the final judgment 
and let $i^{*}_{u,t}$ denote the ground-truth item. A failure case is identified 
by:
\begin{equation}
F_{u,t}
=
\mathbb{I}
\left[
i^{(K_{u,t})}_{u,t}
\neq
i^{*}_{u,t}
\right].
\label{eq:failure_case}
\end{equation}
For each failure, AgentCom constructs a trace containing the complete decision 
trajectory, the selected communication skills, the advisor communication process, and the ground-truth item:
\begin{equation}
\mathcal{T}_{u,t}
=
\left(
\mathcal{Q}_{u,t},
\mathbf{p}_{u,t},
\mathcal{D}_{u,t}, \mathcal{A}_{u,t},
i^{*}_{u,t}
\right),
\label{eq:failure_trace}
\end{equation}
where $\mathcal{D}_{u,t}$ collects the advisor messages and synthesized evidence 
generated throughout communication.

An LLM-based failure analyzer reviews this trace together with the current skill 
bank and produces a structured diagnosis:
\begin{equation}
\left(
\ell^{*}_{u,t},
\kappa_{u,t},
\delta_{u,t}
\right)
=
\operatorname{Analyze}_{\mathrm{LLM}}
\left(
\mathcal{T}_{u,t},
\mathcal{B}_{r}
\right),
\label{eq:failure_diagnosis}
\end{equation}
where $\ell^{*}_{u,t}\in
\{\mathrm{why},\mathrm{what},\mathrm{how},\mathrm{who}\}$ identifies the layer 
responsible for the failure. The variable $\kappa_{u,t}$ specifies the failure 
type, while $\delta_{u,t}$ provides an evidence-grounded explanation that 
describes what was inadequate in the selected path and what communication 
capability was required instead. The analyzer distinguishes three failure types and determines the corresponding
evolution action:
\begin{equation}
\alpha_{u,t}
=
\begin{cases}
\mathrm{SkillUpdate},
& \kappa_{u,t}=\mathrm{skill\ insufficiency},
\\
\mathrm{SkillGeneration},
& \kappa_{u,t}=\mathrm{capability\ gap},
\\
\mathrm{ReRouting},
& \kappa_{u,t}=\mathrm{routing\ mismatch}.
\end{cases}
\label{eq:evolution_action}
\end{equation}

\paragraph{Skill update.}
A skill insufficiency indicates that the selected skill addresses the correct
general communication need but is too broad or incomplete for the failed case.
Let $s^{\mathrm{base}}_{u,t}=s^{\ell^{*}_{u,t}}_{u,t}$ denote the skill selected
at the diagnosed failure layer $\ell^{*}_{u,t}$. Directly modifying this skill
could reduce its usefulness for more general cases. AgentCom therefore applies
an LLM-based refinement operation:
\begin{equation}
s^{\mathrm{child}}
=
\operatorname{Refine}_{\mathrm{LLM}}
\left(
s^{\mathrm{base}}_{u,t},
\delta_{u,t},
\mathcal{T}_{u,t}
\right),
\label{eq:skill_update}
\end{equation}
where $\operatorname{Refine}_{\mathrm{LLM}}$ compares the selected skill
with the diagnosed failure reason $\delta_{u,t}$ and its supporting trace
$\mathcal{T}_{u,t}$. It then produces a specialized skill:
\begin{equation}
s^{\mathrm{child}}
=
\left(
c_{s^{\mathrm{child}}},
a_{s^{\mathrm{child}}},
o_{s^{\mathrm{child}}}
\right),
\label{eq:child_structure}
\end{equation}
where $c_{s^{\mathrm{child}}}$ restricts the skill to the diagnosed failure
pattern, $a_{s^{\mathrm{child}}}$ introduces the additional operation required
to address that pattern, and $o_{s^{\mathrm{child}}}$ specifies the evidence
that should be returned.

The refined skill is added as a child of $s^{\mathrm{base}}_{u,t}$:
\begin{equation}
\operatorname{parent}
\left(
s^{\mathrm{child}}
\right)
=
s^{\mathrm{base}}_{u,t},
\qquad
\mathcal{B}^{\ell^{*}_{u,t}}_{r+1}
=
\mathcal{B}^{\ell^{*}_{u,t}}_{r}
\cup
\left\{
s^{\mathrm{child}}
\right\}.
\label{eq:child_update}
\end{equation}

Here, $\operatorname{parent}(s)$ denotes the immediate parent of $s$ in the
hierarchical bank. Thus, $s^{\mathrm{child}}$ preserves the general purpose of
the selected skill while specializing it for a narrower communication need.
The original skill remains unchanged and can still be used for general cases.

\paragraph{Skill generation.}
A capability gap indicates that the required communication strategy cannot be
obtained by specializing an existing skill. AgentCom therefore applies an
LLM-based generation operation:
\begin{equation}
s^{\mathrm{new}}
=
\operatorname{Generate}_{\mathrm{LLM}}
\left(
\ell^{*}_{u,t},
\delta_{u,t},
\mathcal{T}_{u,t}
\right),
\label{eq:skill_generation}
\end{equation}
where $\operatorname{Generate}_{\mathrm{LLM}}$ constructs a new communication
capability directly from the responsible layer, the diagnosed failure reason,
and the supporting trace. The generated skill is represented as:
\begin{equation}
s^{\mathrm{new}}
=
\left(
c_{s^{\mathrm{new}}},
a_{s^{\mathrm{new}}},
o_{s^{\mathrm{new}}}
\right),
\label{eq:new_skill_structure}
\end{equation}
where its applicability condition, executable instruction, and output contract
are generated to address the previously uncovered communication need.

To place $s^{\mathrm{new}}$ in the hierarchy, AgentCom identifies the most
closely related existing skill in the responsible layer, denoted by
$s^{\mathrm{anchor}}$. The anchor determines only the structural position of
the new skill. Because $s^{\mathrm{new}}$ represents an alternative strategy
rather than a specialization of $s^{\mathrm{anchor}}$, the two skills are placed
under the same parent:
\begin{equation}
\operatorname{parent}
\left(
s^{\mathrm{new}}
\right)
=
\operatorname{parent}
\left(
s^{\mathrm{anchor}}
\right),
\qquad
\mathcal{B}^{\ell^{*}_{u,t}}_{r+1}
=
\mathcal{B}^{\ell^{*}_{u,t}}_{r}
\cup
\left\{
s^{\mathrm{new}}
\right\}.
\label{eq:sibling_generation}
\end{equation}

Therefore, $s^{\mathrm{new}}$ becomes a sibling of
$s^{\mathrm{anchor}}$ and provides an alternative capability at the same layer.
If no related anchor exists, it is inserted as a new top-level skill.

\paragraph{Routing correction.}
A routing mismatch indicates that the public bank already contains a suitable
skill for addressing the diagnosed communication need, but the personalized
router selected an inappropriate skill at layer $\ell^{*}_{u,t}$. In this case,
the failure does not reveal a deficiency in the existing skill definitions and
therefore does not require the bank to be modified or expanded. Instead,
AgentCom uses the failure diagnosis $\delta_{u,t}$ as corrective routing
guidance and reruns personalized skill routing:
\begin{equation}
\widetilde{\mathbf{p}}_{u,t}
=
\operatorname{ReRoute}_{\mathrm{LLM}}
\left(
\mathcal{B}_{r},
\mathcal{X}_{u,t},
\mathbf{p}_{u,t},
\ell^{*}_{u,t},
\delta_{u,t}
\right),
\label{eq:routing_correction}
\end{equation}
where $\mathbf{p}_{u,t}$ is the originally selected communication path and
$\widetilde{\mathbf{p}}_{u,t}$ is the corrected path. Rerouting starts from the
diagnosed layer $\ell^{*}_{u,t}$, replaces the unsuitable skill with a more
appropriate existing skill, and reconstructs all downstream selections because
each subsequent layer is conditioned on the preceding routing decisions. The
corrected path is then executed to obtain new advisor evidence and support a
new UserAgent decision.

After failure analysis, AgentCom either specializes an existing capability
through a child node, introduces an alternative capability through a sibling
node, or corrects an unsuitable route by reusing existing skills. Skill update
and skill generation modify the public bank, whereas routing correction leaves
the bank unchanged. The users involved in the corresponding failure cases are
then rerouted over the resulting bank and re-execute the communication process.
The overall algorithm of AgentCom can be found in
Appendix~\ref{sec:overall_algorithm}.

\section{Experiments}
\label{sec:experiments}









\subsection{Experimental Settings}
\label{sec:experimental_settings}

\subsubsection{Datasets}
\label{sec:datasets}

We conduct experiments on three public datasets with different domains and social structures. \textbf{LastFM}~\cite{cantador2011second} contains artist-listening interactions and user friendship relations. \textbf{Epinions}~\cite{cai2017spmc} contains user--item interactions from a product-review platform together with user trust links. \textbf{LibraryThing}~\cite{zhao2015improving} contains book-related interactions and social connections among readers. These datasets are suitable for our study because they provide both historical preference evidence and inter-user relations that can be used to construct advisor pools. Table~\ref{tab:dataset_statistics} reports their statistics.

\begin{table}[h]
\centering
\small
\caption{Statistics of the experimental datasets.}
\label{tab:dataset_statistics}
\begin{tabular}{lrrrr}
\toprule
Dataset & \#Users & \#Items & \#Interactions & \#Social Links \\
\midrule
LastFM       & 1,877 & 17,612 & 90,924 & 25,218 \\
Epinions     & 1,026 &  9,590 & 11,996 & 15,337 \\
LibraryThing &   860 & 25,251 & 54,306 &  6,187 \\
\bottomrule
\end{tabular}
\end{table}

\subsubsection{Compared Methods}
\label{sec:baselines}

We select baselines from three recommendation paradigms. This design examines whether AgentCom only benefits LLM-based agents or can serve as a general communication layer over different recommendation backbones. The three categories also represent different evidence sources: 

\begin{itemize}[leftmargin=*]
    \item \textbf{Traditional sequential recommendation.} \textbf{SASRec}~\cite{01kang2018self} models a user's ordered interaction history with self-attention and predicts the next item directly. It provides a representative non-social, non-agentic sequential backbone.
\item \textbf{Social recommendation.}
\textbf{GBSR}~\cite{42yang2024graph} incorporates social relations into recommendation, allowing us to compare AgentCom with a model that already uses inter-user connections at the representation level. The comparison tests whether explicit advisor communication offers value beyond social-graph modeling.
\item \textbf{Agentic recommendation.}
We include three agentic recommenders. \textbf{AFL}~\cite{04cai2025agentic} uses an agent feedback loop to evaluate and revise candidate recommendations. \textbf{iAgent}~\cite{02xu2025iagent} formulates recommendation as an interactive agent decision process. \textbf{MemRec}~\cite{03chen2026memrec} further uses explicit interaction memory to support candidate evaluation and refinement. 
\end{itemize}

\subsubsection{Evaluation Metric}
\label{sec:evaluation_metric}

We evaluate recommendation performance using Hit@1, which measures whether the final selected item matches the ground-truth next item. A larger Hit@1 indicates more accurate final recommendation decisions.

\subsubsection{Implementation Details}
\label{sec:implementation_details}

We use deepseek-4-flash~\cite{08guo2025deepseek} as the language model for all LLM-based user and advisor agents. The system is implemented as an agentic reranking framework built on top of a general recommender. Specifically, we first tune SASRec on each dataset and use its best-performing hyperparameter configuration to construct the candidate pool. For each test instance, SASRec generates a 20-item candidate set consisting of the ground-truth item and the 19 highest-ranked non-ground-truth items. As the candidate-generation baseline, SASRec directly takes its top-ranked item as the final recommendation, whereas all other compared methods evaluate or rerank the same 20-item candidate set. Hit@1 is then computed based on the final item selected by each method.

This shared-candidate protocol controls the upstream retrieval space, ensuring that performance differences primarily arise from how each method evaluates and selects among the available candidates. For methods augmented with AgentCom, the backbone first produces a provisional judgment. AgentCom then selects and executes a personalized \textit{why--what--how--who} communication path, aggregates the resulting advisor evidence, and asks the target UserAgent to make the final decision. Most stages of this process are performed through the reasoning capabilities of the underlying language model. Advisor selection additionally incorporates structured signals from the recommendation data. Specifically, social advisors are selected according to the observed social relations provided by the corresponding datasets, while peer similarity is measured using the similarity between user embeddings learned by SASRec. We use identical candidate information, user histories, and language-model configurations across all agentic methods to ensure a fair comparison.

\subsection{Performance Comparison}
\label{sec:main_results}

This experiment examines whether AgentCom can consistently improve recommendation 
accuracy across different backbone models and whether its benefits generalize 
across sequential, social, and agentic recommendation paradigms. For each 
backbone, we compare its original result with the result obtained after adding 
AgentCom, denoted by \textit{+ AgentCom}. Both variants operate on the same 
candidate set. The original backbone provides the initial candidate ordering or 
user judgment, after which AgentCom performs personalized advisor communication 
and allows the target UserAgent to make the final selection. This comparison 
isolates the contribution of AgentCom from that of the underlying recommender. 
Table~\ref{tab:main_results} reports Hit@1 on the three datasets.

\begin{table}[t]
\centering
\small
\caption{Performance comparison in terms of Hit@1. AgentCom-enhanced variants are highlighted in 
blue, with their results shown in bold.}
\label{tab:main_results}
\resizebox{0.98\linewidth}{!}{%
\begin{tabular}{lllccc}
\toprule
Category & Backbone & Variant & LastFM & Epinions & LibraryThing \\
\midrule

\multirow{2}{*}{Sequential}
& \multirow{2}{*}{SASRec}
& Original
& 0.0155 & 0.0010 & 0.0151 \\

& &
\cellcolor{blue!8}\textbf{+ AgentCom}
& \cellcolor{blue!8}\textbf{0.2019}
& \cellcolor{blue!8}\textbf{0.1784}
& \cellcolor{blue!8}\textbf{0.2098} \\

\midrule

\multirow{2}{*}{Social}
& \multirow{2}{*}{GBSR}
& Original
& 0.1630 & 0.1589 & 0.1395 \\

& &
\cellcolor{blue!8}\textbf{+ AgentCom}
& \cellcolor{blue!8}\textbf{0.2375}
& \cellcolor{blue!8}\textbf{0.2273}
& \cellcolor{blue!8}\textbf{0.1930} \\

\midrule

\multirow{6}{*}{Agentic}
& \multirow{2}{*}{AFL}
& Original
& 0.1460 & 0.1140 & 0.1174 \\

& &
\cellcolor{blue!8}\textbf{+ AgentCom}
& \cellcolor{blue!8}\textbf{0.2071}
& \cellcolor{blue!8}\textbf{0.2297}
& \cellcolor{blue!8}\textbf{0.1814} \\

\cmidrule(lr){2-6}

& \multirow{2}{*}{iAgent}
& Original
& 0.2205 & 0.2193 & 0.2023 \\

& &
\cellcolor{blue!8}\textbf{+ AgentCom}
& \cellcolor{blue!8}\textbf{0.2675}
& \cellcolor{blue!8}\textbf{0.2347}
& \cellcolor{blue!8}\textbf{0.2291} \\

\cmidrule(lr){2-6}

& \multirow{2}{*}{MemRec}
& Original
& 0.2648 & 0.2427 & 0.2430 \\

& &
\cellcolor{blue!8}\textbf{+ AgentCom}
& \cellcolor{blue!8}\textbf{0.2696}
& \cellcolor{blue!8}\textbf{0.2485}
& \cellcolor{blue!8}\textbf{0.2634} \\

\bottomrule
\end{tabular}%
}
\end{table}

\begin{itemize}[leftmargin=*]

\item AgentCom improves Hit@1 across different datasets and conventional non-agentic recommenders, showing that UserAgent evaluation and external advisor evidence can substantially revise the initial ranking produced by a sequential model and social recommendation model.

\item AgentCom also consistently improves existing agentic recommenders. AFL 
gains 0.0611, 0.1157, and 0.0640 across the three datasets, while iAgent gains 
0.0470, 0.0154, and 0.0268. Thus, internal agent reasoning and generic feedback 
loops do not fully exploit the complementary evidence available from other 
users. Personalized communication skills provide an additional mechanism for identifying and correcting preference mismatches.

\item MemRec is the strongest original backbone on all three datasets, leaving 
less room for improvement. Nevertheless, AgentCom increases its Hit@1 from 
0.2648 to 0.2696 on LastFM, from 0.2427 to 0.2485 on Epinions, and from 0.2430 
to 0.2634 on LibraryThing, corresponding to relative improvements of 1.8\%, 
2.4\%, and 8.4\%. MemRec with AgentCom achieves the best result on every dataset, demonstrating that cross-user communication complements rather than 
replaces interaction memory.

\end{itemize}

\subsection{Ablation Study}
\label{sec:ablation}

\begin{figure}[h]
    \centering
    \includegraphics[width=\columnwidth]{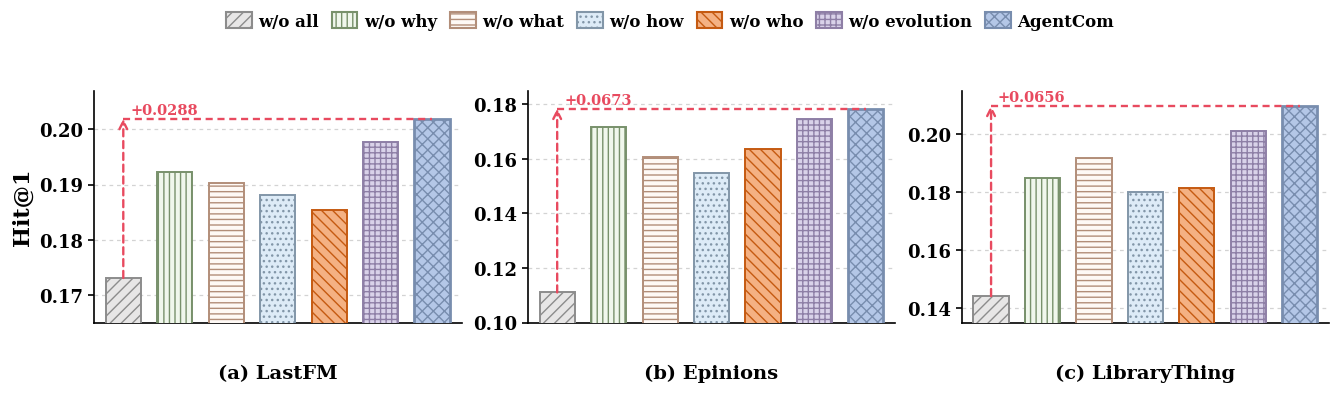}
    \caption{Ablation results of AgentCom on three datasets. The dashed arrows indicate the performance improvements of the complete AgentCom over \textit{w/o all}.}
    \label{fig:ablation}
\end{figure}

We conduct an ablation study to examine how each layer contributes to the communication process. Specifically, we remove the \textit{why}, \textit{what}, \textit{how}, and \textit{who} layers separately. In \textit{w/o why}, the router selects a communication task directly from the current decision state. In \textit{w/o what}, advisors receive no task-specific instruction and comment freely on the candidates. In \textit{w/o how}, the system randomly adopts either single-advisor or multi-advisor communication without selecting a suitable interaction protocol. In \textit{w/o who}, advisors are selected randomly. In \textit{w/o evolution}, the framework retains the initialized communication skill bank but disables failure-driven updates, including the refinement of existing skills and the introduction of new skills. Finally, \textit{w/o all} replaces the entire structured communication skill with generic dialogue. All variants use the same SASRec candidate sets, number of communication rounds, and Hit@1 evaluation setting. As shown in Figure~\ref{fig:ablation}, the complete AgentCom achieves the best performance on all three datasets.

\begin{itemize}[leftmargin=*]

\item Removing any individual layer of the \textit{why--what--how--who} structure
causes a performance decrease, confirming that each layer contributes a distinct
function to the communication process. The \textit{why} layer
identifies the decision deficiency that motivates communication; the
\textit{what} layer translates this deficiency into a focused evidence request;
the \textit{how} layer selects an appropriate protocol for organizing advisor
interactions; and the \textit{who} layer retrieves advisors capable of providing
relevant and reliable evidence. These results show that effective communication
requires not only additional information, but also a structured process for
determining why it is needed, what should be discussed, how it should be
exchanged, and who should provide it.

\item Removing the evolution mechanism keeps the initialized communication skill bank fixed throughout training. Although the predefined skills already provide effective general communication routes, they cannot fully cover user-specific decision failures or communication patterns that are absent from the initial design. The consistent performance decrease across all three datasets shows that failure-driven evolution further improves the framework by refining existing skills and introducing new skills for previously uncovered communication needs.

\item Removing all four layers reduces communication to generic dialogue without 
a structured connection between the decision deficiency, information task, 
interaction protocol, and advisor source. Errors arising at these stages can 
accumulate: the discussion may address the wrong issue, collect scattered 
opinions, and involve advisors whose evidence is irrelevant to the target user. 
Accordingly, \textit{w/o all} performs substantially worse than the complete 
framework, showing that AgentCom benefits from coordinating 
\textit{why--what--how--who} as an integrated communication process rather than 
simply introducing additional dialogue.

\end{itemize}

\subsection{Hyperparameter Analysis}
\label{sec:hyperparameter}

\begin{figure}[H]
    \centering
    \includegraphics[width=\columnwidth]{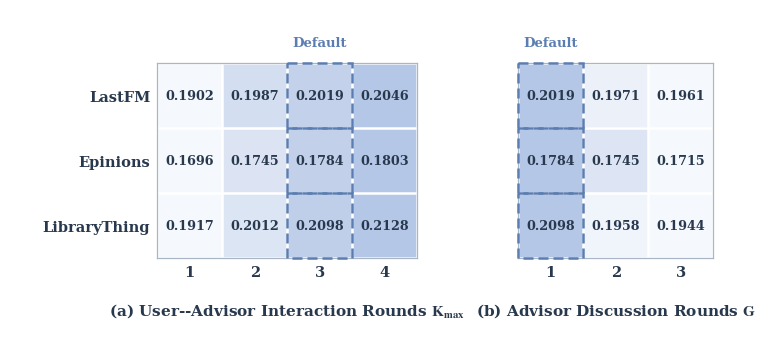}
    \caption{Effect of the maximum number of user--advisor interaction rounds
    $K_{\max}$ and the number of internal advisor discussion rounds $G$.}
    \label{fig:hyperparameter}
\end{figure}

We investigate communication depth through two hyperparameters: the maximum
number of user--advisor interaction rounds $K_{\max}$ and the number of
internal advisor discussion rounds $G$. Here, $K_{\max}$ is the upper bound
on the actual number of user--advisor communication rounds. Experiments are conducted on three datasets with SASRec as the backbone, while the remaining settings are
kept unchanged. The results are presented in
Figure~\ref{fig:hyperparameter}.

\begin{itemize}[leftmargin=*]

\item With the number of internal advisor discussion rounds fixed at $G=1$,
we vary the maximum number of user--advisor interaction rounds $K_{\max}$
from 1 to 4. Increasing $K_{\max}$ improves Hit@1 on all three
datasets, indicating that additional interaction rounds allow the UserAgent
to progressively incorporate advisor evidence and refine its previous decision. However, the marginal gain becomes
smaller after the third round. Increasing $K_{\max}$ from 1 to 3 yields gains
of 0.0117, 0.0088, and 0.0181 on three datasets, whereas the fourth round contributes less. Since each interaction round also requires additional LLM
calls for advisor communication and UserAgent re-decision, we set
$K_{\max}=3$ by default to balance recommendation accuracy and communication cost.

\item With the maximum number of user--advisor interaction rounds fixed at
$K_{\max}=3$, we vary the number of internal advisor discussion rounds $G$
from 1 to 3. Increasing $G$ decreases Hit@1 on all three datasets,
indicating that additional advisor-only discussion does not necessarily provide
more useful evidence for the UserAgent. Increasing $G$ from 1 to 2 causes
absolute performance drops of 0.0048, 0.0039, and 0.0140 on three datasets, while increasing $G$ to 3 enlarges the
corresponding drops to 0.0058, 0.0069, and 0.0154. The degradation is
particularly pronounced on LibraryThing, suggesting that repeated advisor
discussion may introduce redundant information, amplify secondary preference
signals, or shift attention away from the user's current decision needs. Since
additional discussion rounds incur more LLM calls while consistently reducing
recommendation accuracy, we set $G=1$ by default.

\end{itemize}

\subsection{Case Study}
\label{sec:case_study}
We present three representative cases from LibraryThing to illustrate how AgentCom affects the UserAgent's decision, including two successful cases and one unsuccessful case.

\begin{successcasebox}{
Successful Case 1: Correcting Perspective Narrowing\\
(Ground-truth: Black Talk, Prediction: Black Talk)
}
\textbf{User9508, Decision: \textit{Black Talk}. Reason:}
The UserAgent initially favors \textit{The Cage} because of the user's interest in war memoirs. However, the advisors point out that the user's history also contains \casepos{niche non-fiction essays} and works centered on \casepos{social and political commentary}. While \textit{The Cage} mainly matches the war-related theme, \textit{Black Talk} covers \casepos{race, politics, and social identity in an essay-based form}, providing a broader match to the user's reading preferences. The UserAgent therefore \casepos{revises its initial choice and selects \textit{Black Talk}}.
\end{successcasebox}

\noindent
\textbf{Analysis:}
This case succeeds because the initial decision considers only the most obvious preference signal. The advisors recover other relevant evidence from the user's history, giving the UserAgent a more complete view of the user's interests and helping it correct the initial judgment.

\begin{successcasebox}{
Successful Case 2: Clarifying Ambiguous Preference\\
(Ground-truth: Water Elephants, Prediction: Water Elephants)
}
\textbf{User19584, Decision: \textit{Water Elephants}. Reason:}
The UserAgent initially hesitates between \textit{Hashish} and \textit{Water Elephants} under the vague preference for a ``classic'' work. The advisors make this preference more concrete by comparing the candidates in terms of \casepos{classic status, optimistic tone, and narrative appeal}. They find that \textit{Hashish} is more clearly a classic but has an uncertain tone, while \textit{Water Elephants} provides \casepos{stronger evidence of a hopeful and character-driven narrative}. Based on this better-supported evidence, the UserAgent \casepos{selects \textit{Water Elephants}}.
\end{successcasebox}

\noindent
\textbf{Analysis:}
The initial hesitation comes from using the broad label ``classic'' as the main criterion. The advisors decompose this ambiguous preference into several comparable dimensions and avoid treating uncertain attributes as confirmed evidence. This allows the UserAgent to rely on the better-supported narrative and emotional preferences and make the correct judgment.

\begin{failurecasebox}{
Unsuccessful Case: Overemphasis on Secondary Preferences\\
(Ground-truth: Passion, Prediction: Nightshade)
}
\textbf{User25209, Decision: \textit{Nightshade}. Reason:}
The UserAgent initially selects the correct item \casepos{\textit{Passion}}. The advisors identify that \casepos{\textit{Passion} has a stronger romantic subplot}, but they also emphasize that \textit{Nightshade} offers \caseneg{stronger suspense, a darker atmosphere, and easier series entry}. As these secondary advantages of \textit{Nightshade} receive \caseneg{increasing attention in later communication}, they gradually outweigh the stronger romantic evidence supporting \textit{Passion}. The UserAgent consequently \caseneg{overturns its initial correct decision and selects \textit{Nightshade}}.
\end{failurecasebox}

\noindent
\textbf{Analysis:}
This case fails because the communication repeatedly reinforces the suspense, dark atmosphere, and accessibility of \textit{Nightshade}, although these are secondary to the user's stronger romantic preference. The repeated emphasis distorts the relative importance of the available evidence and causes the UserAgent to replace an initially correct choice with an incorrect one.

\section{Related Work}
\label{sec:related_work}

Traditional recommender systems learn user preferences from historical 
interactions and directly rank candidate items. Sequential models capture temporal dependencies in user behavior~\cite{01kang2018self,30sun2019bert4rec}, while 
graph-based collaborative filtering methods propagate 
preference signals over user--item graphs~\cite{10he2020lightgcn}. 
Despite their increasingly powerful representations, these methods generally 
follow a one-shot prediction paradigm and provide no user-side mechanism for examining or revising a recommendation after it is generated. 

The emergence of LLMs has shifted recommendation toward agentic systems. Early 
studies primarily employed LLM-powered agents to construct user profiles and 
simulate user behaviors, supporting interactive evaluation and environment 
modeling~\cite{11zhang2024generative,27wang2025user}. Other works equipped 
recommendation agents with planning, tool use, and autonomous reasoning 
capabilities~\cite{28wang2024recmind,25zhao2024let}. These developments gradually 
extended the UserAgent from a passive behavior simulator to an active participant 
that evaluates recommendations and provides natural-language feedback.
Recent methods further integrate UserAgents into the recommendation process. AFL 
constructs an iterative feedback loop in which the recommendation agent proposes 
items and the UserAgent evaluates them, jointly improving recommendation and user 
simulation~\cite{04cai2025agentic}. iAgent positions a user-side agent as a shield 
between the user and the platform, combining user instructions, external knowledge, 
self-reflection, and dynamic memory to rerank recommendations according to individual 
interests~\cite{02xu2025iagent}. MemRec introduces collaborative memory and 
extracts high-signal context from a user--item memory graph 
for downstream recommendation reasoning~\cite{03chen2026memrec}. Although these 
methods improve agentic recommendation through feedback, protection, or memory, 
they do not formulate communication among UserAgents as reusable 
skills. AgentCom addresses this gap by routing and evolving advisor communication 
for each target user.

\section{Conclusion}
\label{sec:conclusion}

In this work, we introduced AgentCom, a personalized communication skill framework for agentic recommendation. AgentCom addresses perspective narrowing by allowing a target UserAgent to obtain complementary evidence from other users before making its final decision. Specifically, AgentCom organizes reusable capabilities into a shared \textit{why--what--how--who} skill bank. A personalized router determines why communication is needed, what evidence should be requested, how advisors should interact, and who can execute the selected protocol. A failure-driven evolution mechanism further attributes unsuccessful outcomes and accordingly refines existing public skills or expands the shared skill bank.

In the future, we will investigate how to make agent communication more efficient and cost-effective by reducing unnecessary interaction rounds, prioritizing high-value evidence, and balancing communication cost against expected recommendation gains. We will also explore whether complementary evidence can be obtained without direct communication among UserAgents, such as through reusable shared knowledge or compact preference representations. 
Finally, we will examine the privacy and security risks of cross-user communication, including unintended preference disclosure, unreliable or malicious advisors, and the propagation of biased or manipulated evidence.

\bibliographystyle{ACM-Reference-Format}
\bibliography{Agentcom}

\clearpage
\appendix

\section{Theory-Guided Skill Bank Initialization}
\label{app:skill_initialization}

Table~\ref{tab:initial_skill_bank} lists the public nodes used to initialize AgentCom before failure-driven evolution. The layers follow the same execution order as the personalized router: \textit{why--what--how--who}.

\begin{table}[h]
\centering
\small
\caption{Theory-guided initialization of the public communication skill bank.}
\label{tab:initial_skill_bank}

\begin{tabular}{
    p{0.05\textwidth}
    p{0.35\textwidth}
}
\toprule
Layer & Initial skills \\
\midrule

Why
&
Cold start~\cite{schein2002methods}; candidate conflict~\cite{40tversky1992choice,bollen2010understanding}; novelty uncertainty~\cite{wood2006fear,loewenstein1994psychology}; internal--prior conflict~\cite{morvan2017analysis,larrick2004debiasing}.
\\

What
&
Reduce the hesitation set~\cite{bollen2010understanding,hauser1990evaluation}; find an interested subset~\cite{hauser1990evaluation}; compare remaining candidates~\cite{payne1993adaptive,hauser1990evaluation}; check the user's reasoning~\cite{larrick2004debiasing}.
\\

How
&
Single advisor~\cite{34yaniv2004receiving}; multi-advisor cooperation~\cite{mesmer2009information}; multi-advisor competition~\cite{nemeth1996dissent}.
\\

Who
&
Trusted users~\cite{39ma2009learning,sniezek2001trust}; similar users~\cite{resnick1994grouplens,sniezek2001trust}; experienced users~\cite{sniezek2001trust}; friend-of-friend advisors~\cite{granovetter1973strength}.
\\

\bottomrule
\end{tabular}
\end{table}

\section{Overall Algorithm}
\label{sec:overall_algorithm}

Algorithm~\ref{alg:agentcom} summarizes the training procedure of AgentCom. Starting from the theory-guided skill bank $\mathcal{B}_0$, each evolution
round applies personalized routing and communication execution to the training
interactions and collects unsuccessful cases. For each failure, the analyzer identifies the responsible layer $\ell^{*}_{u,t}$, failure type
$\kappa_{u,t}$, and explanation $\delta_{u,t}$. A skill insufficiency is addressed by adding a specialized child, a capability gap by introducing an alternative sibling, and a routing mismatch by selecting a more appropriate existing skill without modifying the bank. The affected interaction is then rerouted from the diagnosed layer and re-executed. After $R$ evolution rounds, the resulting skill bank is fixed for inference.

\begin{algorithm}[h]
\small
\caption{Overall Training Algorithm of AgentCom}
\label{alg:agentcom}

\KwIn{
Training interactions $\mathcal{D}_{\mathrm{tr}}$,
user histories $\{h_u\}$,
candidate sets $\{\mathcal{C}_{u,t}\}$,
communication-round counts $\{K_{u,t}\}$,
and maximum evolution rounds $R$
}

\KwOut{
Evolved communication skill bank $\mathcal{B}_R$
}

Initialize the theory-guided skill bank $\mathcal{B}_0$\;

\For{$r\leftarrow 0$ \KwTo $R-1$}{

    Initialize the failure set
    $\mathcal{F}_r\leftarrow\emptyset$\;

    Set
    $\mathcal{B}_{r+1}\leftarrow\mathcal{B}_r$\;

    \ForEach{$
    (u,t,\mathcal{C}_{u,t},i^*_{u,t})
    \in\mathcal{D}_{\mathrm{tr}}
    $}{

        Obtain the initial judgment
        $\mathbf{q}^{(0)}_{u,t}$
        according to Eq.~\eqref{eq:provisional_user_judgment}\;

        Construct the routing context
        $\mathcal{X}_{u,t}$\;

        Route over $\mathcal{B}_r$ to obtain $\mathbf{p}_{u,t}$\;

        Execute $\mathbf{p}_{u,t}$ for $K_{u,t}$ communication rounds
        to obtain the communication trace $\mathcal{D}_{u,t}$,
        evidence packet $\mathcal{A}_{u,t}$, and decision trajectory
        $\mathcal{Q}_{u,t}$\;

        Extract the final choice
        $i^{(K_{u,t})}_{u,t}$
        from $\mathbf{q}^{(K_{u,t})}_{u,t}$\;

        \If{$
        i^{(K_{u,t})}_{u,t}\neq i^*_{u,t}
        $}{

            Construct the failure trace
            $\mathcal{T}_{u,t}$ according to
            Eq.~\eqref{eq:failure_trace}\;

            Add
            $(u,t,\mathcal{X}_{u,t},
            \mathbf{p}_{u,t},\mathcal{T}_{u,t})$
            to $\mathcal{F}_r$\;
        }
    }

    \ForEach{$
    (u,t,\mathcal{X}_{u,t},
    \mathbf{p}_{u,t},\mathcal{T}_{u,t})
    \in\mathcal{F}_r
    $}{

        Obtain
        $(\ell^*_{u,t},\kappa_{u,t},\delta_{u,t})$
        through failure analysis\;

        Determine the action
        $\alpha_{u,t}$ according to
        Eq.~\eqref{eq:evolution_action}\;

        \uIf{$
        \alpha_{u,t}=\mathrm{SkillUpdate}
        $}{

            Set
            $s^{\mathrm{base}}_{u,t}
            \leftarrow
            s^{\ell^*_{u,t}}_{u,t}$\;

            Generate
            $s^{\mathrm{child}}$ according to
            Eq.~\eqref{eq:skill_update}\;

            Set
            $\operatorname{parent}(s^{\mathrm{child}})
            \leftarrow s^{\mathrm{base}}_{u,t}$\, and update the responsible sub-bank;

            Reroute from layer $\ell^*_{u,t}$ over
            $\mathcal{B}_{r+1}$ to obtain
            $\widetilde{\mathbf{p}}_{u,t}$\;
        }

        \uElseIf{$
        \alpha_{u,t}=\mathrm{SkillGeneration}
        $}{

            Generate $s^{\mathrm{new}}$ according to
            Eq.~\eqref{eq:skill_generation}\;

            Place $s^{\mathrm{new}}$
            according to Eq.~\eqref{eq:sibling_generation}\,and update the responsible sub-bank;

            Reroute from layer $\ell^*_{u,t}$ over
            $\mathcal{B}_{r+1}$ to obtain
            $\widetilde{\mathbf{p}}_{u,t}$\;
        }

        \Else{

            Obtain the corrected path
            $\widetilde{\mathbf{p}}_{u,t}$
            according to Eq.~\eqref{eq:routing_correction}\;
        }

        Re-execute $\widetilde{\mathbf{p}}_{u,t}$ for
        $K_{u,t}$ communication rounds to obtain the corrected
        UserAgent decision\;
    }
}

\KwRet{$\mathcal{B}_R$}\;

\end{algorithm}

\section{LLM Cost Analysis Experiments}
\label{sec:token_consumption}

In this section, we quantify the LLM inference overhead introduced by AgentCom and examine whether it is justified by the resulting recommendation gains. All variants are evaluated on the same test users and candidate sets, and all LLM-based components use the same language-model configuration. Table~\ref{tab:api_cost} reports the estimated API cost per user. 
The zero values for SASRec and GBSR indicate that their original recommendation procedures do not invoke an LLM. Therefore, the reported values reflect LLM usage rather than the total computational cost of the recommender.

Although AgentCom incurs additional LLM API cost for each user, the resulting overhead remains practically acceptable for two reasons. First, under the adopted API pricing, the additional cost is less than \$0.01 per user and varies only marginally across different backbones, making it modest and predictable. Second, the communication overhead is accompanied by substantial improvements in recommendation performance, especially for lightweight and non-agentic recommenders. 
These results suggest that the LLM inference cost is well justified by the corresponding accuracy gains, although the performance--cost trade-off is more favorable for lightweight backbones than for already strong agentic recommenders.

\begin{table}[h]
\centering
\small
\caption{Average LLM API cost per user in USD. AgentCom-enhanced variants are highlighted in blue, with their costs shown in bold.}
\label{tab:api_cost}
\resizebox{0.98\linewidth}{!}{%
\begin{tabular}{lllccc}
\toprule
Category & Backbone & Variant & LastFM & Epinions & LibraryThing \\
\midrule

\multirow{2}{*}{Sequential}
& \multirow{2}{*}{SASRec}
& Original
& 0.0000 & 0.0000 & 0.0000 \\

& &
\cellcolor{blue!8}\textbf{+ AgentCom}
& \cellcolor{blue!8}\textbf{0.0033}
& \cellcolor{blue!8}\textbf{0.0037}
& \cellcolor{blue!8}\textbf{0.0039} \\

\midrule

\multirow{2}{*}{Social}
& \multirow{2}{*}{GBSR}
& Original
& 0.0000 & 0.0000 & 0.0000 \\

& &
\cellcolor{blue!8}\textbf{+ AgentCom}
& \cellcolor{blue!8}\textbf{0.0037}
& \cellcolor{blue!8}\textbf{0.0040}
& \cellcolor{blue!8}\textbf{0.0037} \\

\midrule

\multirow{6}{*}{Agentic}
& \multirow{2}{*}{AFL}
& Original
& 0.0025 & 0.0027 & 0.0034 \\

& &
\cellcolor{blue!8}\textbf{+ AgentCom}
& \cellcolor{blue!8}\textbf{0.0060}
& \cellcolor{blue!8}\textbf{0.0065}
& \cellcolor{blue!8}\textbf{0.0072} \\

\cmidrule(lr){2-6}

& \multirow{2}{*}{iAgent}
& Original
& 0.0010 & 0.0008 & 0.0010 \\

& &
\cellcolor{blue!8}\textbf{+ AgentCom}
& \cellcolor{blue!8}\textbf{0.0048}
& \cellcolor{blue!8}\textbf{0.0049}
& \cellcolor{blue!8}\textbf{0.0049} \\

\cmidrule(lr){2-6}

& \multirow{2}{*}{MemRec}
& Original
& 0.0038 & 0.0040 & 0.0042 \\

& &
\cellcolor{blue!8}\textbf{+ AgentCom}
& \cellcolor{blue!8}\textbf{0.0077}
& \cellcolor{blue!8}\textbf{0.0083}
& \cellcolor{blue!8}\textbf{0.0083} \\

\bottomrule
\end{tabular}%
}
\end{table}
\end{document}